\documentclass[10pt, twocolumn]{article}
\usepackage{cite}
\usepackage{amssymb,amsfonts}
\usepackage{amsthm}
\usepackage[cmex10]{amsmath} 
\usepackage{algorithmic}
\usepackage{graphicx}
\usepackage{textcomp}
\usepackage{xcolor}
\usepackage{balance} 
\usepackage{url}
\providecommand{\keywords}[1]
{
  \small	
  \textbf{\textit{Keywords---}} #1
}
\newtheorem{theorem}{Theorem}
\newtheorem{corollary}{Corollary}[theorem]

\newtheorem{definition}{Definition}
\newtheorem{remark}{Remark}
\newcommand{\ou}{%
  \mathrel{%
    \vcenter{\offinterlineskip
      \ialign{##\cr$<$\cr\noalign{\kern-1.5pt}$>$\cr}
			 }%
  }%
}	
\def\BibTeX{{\rm B\kern-.05em{\sc i\kern-.025em b}\kern-.08em
    T\kern-.1667em\lower.7ex\hbox{E}\kern-.125emX}}
\begin{document}

\title{Adversarial Classification under Gaussian Mechanism: Calibrating the Attack to Sensitivity
\footnote{This work has been supported by the 3IA C\^ote d'Azur Interdisciplinary Institute for Artificial Intelligence project
with the reference number ANR-19-P3IA-0002. }}

\author{%
Ay\c{s}e \"{U}nsal \qquad Melek \"{O}nen\\ EURECOM, France \\ \url{firstname.lastname@eurecom.fr}
}

\maketitle

\begin{abstract}
  This work studies anomaly detection under differential privacy (DP) with Gaussian perturbation using both statistical and information-theoretic tools. In our setting, the adversary aims to modify the content of a statistical dataset by inserting additional data without being detected by using the DP guarantee to her own benefit.
To this end, we characterize information-theoretic and statistical thresholds for the first and second-order statistics of the adversary's attack, which balances the privacy budget and the impact of the attack in order to remain undetected.
Additionally, we introduce a new privacy metric based on Chernoff information for classifying adversaries under differential privacy as a stronger alternative to $(\epsilon, \delta)-$ and Kullback-Leibler DP for the Gaussian mechanism. Analytical results are supported by numerical evaluations. 

\end{abstract}


\section{Introduction \label{sec:intro}}

The major issue in terms of data privacy in today's world stems from the fact that machine learning (ML) algorithms strongly depend on the use of large datasets to work efficiently and accurately. Along with the highly increased deployment of ML, its privacy aspect rightfully became a cause of concern, since the collection of such large datasets makes users vulnerable to fraudulent use of personal, (possibly) sensitive information. This vulnerability is aimed to be mitigated by privacy enhancing technologies that are designed to protect data privacy of users. 

\textit{Differential privacy} has been proposed to address this problem and it has furthermore been used to develop practical methods for protecting private user-data. Dwork's original definition of DP in \cite{D06} emanates from a notion of statistical indistinguishability of two different probability distributions which is obtained through randomization of the data prior to its publication. The outputs of two DP mechanisms are indistinguishable for two datasets that only differ in one user's data, such datasets are called neighbors. In other words, DP guarantees that the output of the mechanism is \textit{statistically indifferent} to changes made in a single row of the dataset proportional to its privacy budget.

Adversarial classification/anomaly detection is an application of the supervised ML approach, statistical classification, to detect misclassification attacks where adversaries who are aware of the DP mechanism and shield themselves by using it to their benefit. In other words, privacy protection methods are \textit{weaponized} by adversaries in order to avoid being detected. This paper studies anomaly detection in Gaussian mechanisms to establish the trade-off between the impact of the attack and the privacy protection to remain indistinguishable by employing both statistical and information-theoretic tools. In our model, the adversary is aware of the underlying DP mechanism and its parameters and wants to benefit from it using it as an attack tool \cite{r1, r12}. In our setting, we consider an adversary who both aims to discover and harm the information of a dataset by inserting additional data, so the adversary's goal is to maximize the possible damage while minimizing the probability of being detected. Accordingly, we establish stochastic and information-theoretic relations between the impact of the adversary's attack and the privacy budget of the Gaussian mechanism. 

This work, in part, is an extension of \cite{UO21} to Gaussian mechanisms which introduced statistical thresholds of detection in Laplace mechanisms. As for the methodology, in this work, we introduce the Gaussian privacy-distortion function for adversarial classification by deriving the mutual information between the datasets before and after the attack (considered as neighbors). This results in an upper bound on the second-order statistics of the data added to the system by the adversary, in order to determine an information-theoretic threshold for correctly detecting the attack. Originally, the lossy source-coding approach in the information-theoretic DP literature has mostly been used
to quantify the privacy guarantee \cite{WYZ16} or the leakage \cite{SS14,CF12} contrary to the current paper. \cite{PG21} stands out in the way that the rate-distortion perspective is applied to DP where various fidelity criteria is set to determine how fast the empirical distribution converges to the actual source distribution. Alternatively, in \cite{r1}, the authors formulate the problem of the two conflicting goals of the adversary that are maximizing the damage to the DP mechanism while remaining undetected as a multi-criteria optimization problem. In \cite{UO21}, the authors presented an application of the Kullback-Leibler (KL)-DP of \cite{CY16} for detecting misclassification attacks in Laplace mechanisms, where the corresponding distributions in relative entropy were considered as the DP noise with and without the adversary's advantage. This work introduces a novel DP metric based on Chernoff information along with its application to adversarial classification as a stronger alternative to privacy metrics $(\epsilon, \delta)-$ DP and KL-DP for Gaussian mechanism. 

 \paragraph*{Outline}In the upcoming section, we remind the reader of some important preliminaries from the DP literature along with the detailed problem definition and performance criteria. In Section \ref{sec:adv_class}, we present statistical and information-theoretic thresholds for anomaly detection which is followed by Section \ref{sec:chernoff} where we introduce a new metric of DP based on Chernoff information. We present numerical evaluation results in Section \ref{sec:numeric} and draw our final conclusions in Section \ref{sec:conc}.

\section{Preliminaries, Model and Performance Criteria}

Before presenting the addressed problem in detail, the reader is reminded of some preliminaries on DP.
\subsection{Preliminaries for DP \label{subsec:preliminaries}}

Two datasets $\mathbf{X}$ and $\tilde{\mathbf{X}}$ are called neighbors if $d(\mathbf{X},\tilde{\mathbf{X}})=1$ where $d(.,.)$ denotes the Hamming distance.
Accordingly, $(\epsilon, \delta)-$DP is defined by \cite{DR05} as follows.
\begin{definition}\label{def:dp}
A randomized algorithm $\mathcal{M}$ guarantees $(\epsilon, \delta)-$ DP if $\forall \; \mathbf{X},\tilde{\mathbf{X}}$ that are neighbors within the domain of $\mathcal{M}$ and $\forall S \subseteq Range(\mathcal{M})$ the following inequality holds.
\begin{equation}\label{ineq:dp}
\Pr\left[\mathcal{M}(\mathbf{X}) \in S\right] \leq \Pr\left[\mathcal{M}(\tilde{\mathbf{X}}) \in S \right] \exp\{\epsilon\}+\delta
\end{equation}
\end{definition} We will refer to the parameters $\epsilon$ and $\delta$ as \textbf{privacy budget} throughout the paper. Next definition reminds the reader of the $L_2$ norm global sensitivity. 
\begin{definition}[$L_2$ norm sensitivity] \label{def:sens_L2}
$L_2$ norm sensitivity denoted by $s$ refers to the smallest possible upper bound on the $L_2$ distance between the images of a query $q: D \rightarrow \mathbb{R}^k$ when applied to two neighboring datasets $\mathbf{X}$ and $\tilde{\mathbf{X}}$ as
\begin{equation}
s=\sup_{d(\mathbf{X},\tilde{\mathbf{X}})=1}||q(\mathbf{X})-q(\tilde{\mathbf{X}})||_2.
\end{equation}
\end{definition}
Application of Gaussian noise results in a more relaxed privacy
guarantee, that is $(\epsilon, \delta)-$DP contrary to Laplace mechanism, which brings about $(\epsilon, 0)-$ DP. 
$(\epsilon, \delta)-$DP is achieved by calibrating the noise variance as a function of the privacy budget and query sensitivity as given by the next definition.
\begin{definition}\label{def:gaussian_mec}
Gaussian mechanism \cite{DMNA06} is defined for a function (or a query) $q: D \rightarrow \mathbb{R}^k$ as follows
\begin{equation}
\mathcal{M}(\mathbf{X}, q(.), \epsilon, \delta)= q(\mathbf{X})+(Z_1, \cdots, Z_k)
\end{equation} where $Z_i \sim \mathcal{N}(0, \sigma^2)$, $i=1,\cdots, k$ denote independent and identically distributed (i.i.d.) Gaussian random variables with variance $\sigma_z^2= \frac{2 s^2 \log (1.25/\delta)}{\epsilon^2}$.
\end{definition}
\begin{theorem}[\cite{DR05}] For any $\epsilon, \delta \in(0,1)$, the Gaussian mechanism satisfies $(\varepsilon,\delta)$-differential privacy.
\end{theorem}
Lastly, we revisit the so-called Kullback-Leibler (KL) DP definition of \cite{CY16}.
\begin{definition}[KL-DP] \label{def:KL-DP}
For a randomized mechanism $P_{Y|X}$ that guarantees $\epsilon-$ KL-DP, the following inequality holds for all its neighboring datasets $X$ and $\tilde{X}$.
\begin{equation} \label{eq:KLdp}
D(P_{Y|X}||P_{Y|\tilde{X}}) \leq \exp\{\epsilon\}
\end{equation}
\end{definition}

\subsection{Adversarial Classification under DP\label{subsec:hyp}}
We define the original dataset in the following form $\mathbf{X}=X^n=\{X_1, \cdots, X_n\}$. The query function takes the aggregation of this dataset as $q(\mathbf{X})=\sum_i^n X_i$ and the DP-mechanism adds Gaussian noise $Z$ on the query output leading to the noisy output in the following form $\mathcal{M}(\mathbf{X}, q(.), \epsilon, \delta)= Y = \sum_i^n X_i +Z$. This public information is altered by an adversary, who adds a single record denoted $X_a$ to this dataset. The modified output of the DP-mechanism becomes $\sum_i^n X_i + X_a +Z$. 

\subsubsection{First-order statistics of $X_a$ \label{subsec:model_first}} In our first approach, we employ hypothesis testing in a similar vein to \cite{r3} to determine whether or not the defender fails to detect the attack. Accordingly, we set the following hypotheses where the null and alternative hypotheses are respectively translated into DP noise distribution with and without the bias induced by the attacker. 
\begin{equation}
\begin{aligned}\label{eq:ht}
H_{0} &: \textrm{defender\;fails\;to\;detect}\; X_a  \\
H_{1} &: \textrm{defender\;detects}\; X_a 
\end{aligned}
\end{equation} 
False alarm refers to the event when the defender detects the attack when in fact there was no attack with the corresponding probability denoted by $\alpha$. Similarly, mis-detection is failing to detect an actual attack with the probability of occurrence denoted by $\beta$. This second part seeks a trade-off between the shift due to the first-order statistics of the additional adversarial data, the privacy budget, the sensitivity of the query and the probability of false alarm by using the following likelihood ratio function
${\Lambda = \frac{\mathcal{L}(p_1)}{\mathcal{L}(p_0)} \underset{H_1}{\overset{H_0}{\ou}} k}$
where $p_1$ and $p_0$ denote the noise distributions for the $i^{th}$ hypothesis with the corresponding location parameter $\mu_i$ for $i=0,1$.
The impact of the attack for this approach is denoted by $\Delta \mu=\mu_1-\mu_0$.
\subsubsection{Second-order statistics of $X_a$ \label{subsec:model_second}} Our second approach is inspired by rate-distortion theory. We employ the biggest possible difference between the images of the query for the datasets with and without the additional data $X_a$ (i.e. neighbors) as the fidelity criterion by using Definition \ref{def:sens_L2}. Here the traditional notion of distortion is replaced by the sensitivity of the query function. Accordingly, we derive the mutual information between the original dataset and its neighbor in order to bound the additional data's second-order statistics. The goal is to calibrate the standard deviation of the additional data to the original data in order to avoid detection, while giving as much harm as possible to the system. 
To simplify our derivations, we assume that the original dataset $X^{n}=\{X_1, X_2, \cdots,X_i, \cdots, X_n\}$ and its neighbor $\tilde{X}^{n}=\{X_1, X_2, \cdots, X_i, \cdots, X_n+X_a\}$ have the same dimension $n$, where $X_i$ are assumed to be i.i.d following the Gaussian distribution with parameters $\mathcal{N}(0, \sigma^2_{X_i})$. 
\begin{definition} We define the distortion $D$ in adversarial classification under DP as follows
\begin{equation}
D=\sup _{d(\mathbf{X},\tilde{\mathbf{X}})=1} \mathrm{dist}(q(\mathbf{X})-q(\tilde{\mathbf{X}})).
\end{equation}
\end{definition} We employ the squared-error distortion for $\mathrm{dist}(.,.)$, and thus, global sensitivity of Definition \ref{def:sens_L2} is used. 
\begin{definition}\label{def:privacy_dist_func}
The privacy-distortion function $P(D)$ is defined by 
\begin{equation}
P(D)=\underset{f(\tilde{x}|x):\mathbb{E}[\mathrm{dist}(q(\mathbf{X})-q(\tilde{\mathbf{X}}))]\leq D}{\min} I(\mathbf{X}; \tilde{\mathbf{X}})
\end{equation}
\end{definition}
Most related reference \cite{WYZ16}, in fact, defines the \textit{distortion-privacy} problem (contrary to Definition \ref{def:privacy_dist_func}) under different notions of privacy measures, where the Hamming distance-based average distortion is minimized subject to DP, mutual information and identifiability. Furthermore, the distortion in \cite{WYZ16} is used to determine the number of rows between two neighboring datasets that differ. In our model, the adversary aims to calibrate the impact of the attack, more precisely the variance of $X_a$ denoted $\sigma^2_{X_a}$, according to the sensitivity of the DP mechanism.

\section{Adversarial Classification  \label{sec:adv_class}}
In this part, we apply a source-coding approach to anomaly detection under DP, which results in an upper bound on the variance of the additional data as a function of the sensitivity of the mechanism and the original data's statistics by deriving the mutual information between the neighboring datasets. 
Additionally, we present a statistical trade-off between the probability of false alarm, privacy budget and the impact of the attack for the first-order statistics of the data. 

\subsection{Privacy-Distortion Trade-off for Second-Order Statistics \label{subsec:info}}

The idea in this part is to render the problem of adversarial classification under DP as a lossy source-coding problem. Instead of using the mutual information between the input and output (or the input's estimate obtained by using the output) of the mechanism, for this problem we derive the mutual information between the (neighboring) datasets before and after the attack, according to the adversary's conflicting goals as maximizing the induced bias while remaining undetected.
We present our first main result by the following theorem.
\begin{theorem}
The privacy-distortion function for a dataset $X^n$ and Gaussian mechanism as defined by (\ref{def:gaussian_mec}) is
\begin{equation}
P(s)=\frac{1}{2} \log\left( f_n \left( 1+\prod_i^n \sigma_{X_i}^{2}/s^2 \right)\right),
\end{equation} for $s\in \left[0, \prod_i^n \sigma_{X_i}^{2}\right]$ and zero elsewhere. $\sigma_{X_i}$ denotes the standard deviation of $X_i$ for $i=1,\cdots,n$, $f_n$ is some constant dependent on the size of the dataset $n$ and $\sigma_{X_i}$ is the standard deviation of the additional data.
\end{theorem} 
\begin{proof}
The first expansion of $I(X^n;\tilde{X}^{n})$ proceeds as follows 
\begin{align}
&I(X^n;\tilde{X}^{n})= h(X^n)-h(X^n|\tilde{X}^{n})\\
&\geq h(X^n) -h(q(X^n)-\tilde{X}^{n}|\tilde{X}^{n})\label{eq1} \\
&=h(X^n) -h(q(X^n)-q(\tilde{X}^{n})|\tilde{X}^{n}) \\
&\geq h(X^n) -h(q(X^n)-q(\tilde{X}^{n})) \label{eq2}\\
&\geq \frac{1}{2}  \sum_{i=1}^n\log \left(\left(2 \pi \mathrm{e}\right) \sigma^2_{X_i}\right) -\frac{1}{2} \log \left(2 \pi \mathrm{e} s^2\right) \label{eq3} \\
&= \frac{1}{2} \log \left((2 \pi \mathrm{e})^{n-1}\prod_i^n \sigma^{2}_{X_i}/s^2\right) \label{eq:2nd_expansion}
\end{align} In (\ref{eq2}), we apply the following property due to concavity of entropy function, $h(g(x))\leq h(x)$ for any function $g(.)$ and introduce the lower bound since the condition conditioning reduces entropy. In (\ref{eq3}), we plug in Definition \ref{def:sens_L2} into the second term after bounding it by Gaussian entropy.
\end{proof} It is worth noting that the additional factor $2\pi \mathrm{e}$ appears here as opposed to the original rate-distortion function due to the choice of the query function that aggregates the entire dataset and returns an output of size 1.
\begin{corollary} \label{cor:upper_bound}
The second order statistics of the additional data inserted into the dataset by the adversary is upper bounded by a function of the privacy budget $(\epsilon, \delta)-$ and the statistics of the original dataset as follows
\begin{equation}\label{upper_bound_attack}
\sigma^2_{X_a} \leq \frac{1}{(2\pi \mathrm{e})^{n-1}}\left[\frac{s^2}{1-s^2/\sigma^2_{X_n}}\right] 
\end{equation} where $s^2= \frac{\sigma_z^2 \epsilon^2}{2 \log(1.25/ \delta)}$ due to Definition \ref{def:gaussian_mec} for $n\geq 2$.
\end{corollary}
\begin{proof}
For the second expansion of $I(X^n;\tilde{X}^{n})$, we have the following considering the neighbor that includes $X_a$ has now $(n+1)$ entries over $n$ rows as $\tilde{X}^{n}=\{X_1, X_2, \cdots, X_n+X_a\}$.
\begin{align}
&I(X^n;\tilde{X}^{n})= h(\tilde{X}^{n}) - h(\tilde{X}^{n}|X^n)\\
& \leq  \sum_{i=1}^{n}\frac{1}{2} \log \left(2 \pi \mathrm{e}\right)^n \sigma^2_{X_i}-\frac{1}{2} \log \left((2 \pi \mathrm{e})^n \sigma_{X_a}^{2}\right)\label{eq11} \\
&= \frac{1}{2} \log \left((2 \pi \mathrm{e})^n \prod_{i=1}^{n-1} \sigma_{X_i}^{2}(\sigma^2_{X_n}+\sigma_{X_a}^{2})\right) \\
&-\frac{1}{2} \log \left((2 \pi \mathrm{e})^n \sigma_{X_a}^{2}\right)\\
&=\frac{1}{2} \log \prod_{i=1}^{n-1} \sigma^2_{X_i}\left( 1+\frac{\sigma_{X_n}^{2}}{\sigma_{X_a}^{2}}\right) \label{eq:1st_expansion}
\end{align} Due to the adversary's attack, in the first term of (\ref{eq11}), we add up the variances of $(n+1)$ $X_i$'s including $X_a$.
Since (\ref{eq:1st_expansion}) $\geq$ (\ref{eq:2nd_expansion}), global sensitivity is bounded as follows in terms of the second-order statistics of the original data and those of the additional data $X_a$.
\begin{equation}
s \geq (2 \pi \mathrm{e})^{\frac{n-1}{2}}\frac{\sigma_{X_n} \cdot \sigma_{X_a} }{ \left(\sigma^{2}_{X_n} +\sigma^2_{X_a}\right)^{1/2} }
\end{equation} 
Alternatively, the lower bound on the sensitivity of the Gaussian mechanism can be used as an upper bound on $\sigma^2_{X_a}$ to yield a threshold in terms of the additional data's variance as a function of the privacy budget and the original data's statistics to guarantee that the adversary avoids being detected. \end{proof}
\begin{remark}
The second expansion of the mutual information between neighboring datasets derived in (\ref{eq:2nd_expansion}), can be related to the well-known \textbf{rate-distortion function of the Gaussian source} which, originally, provides the minimum possible transmission rate for a given distortion balancing (mostly for the Gaussian case) the squared-error distortion with the source variance. This is in line with the adversary's goal in our setting, where the adversary aims to maximize the damage that s/he inflicts on the DP-mechanism. But at the same time to avoid being detected, the attack is calibrated according to the sensitivity which here replaces the distortion. Thus, similar to the classical rate-distortion theory, here the mutual information between the neighbors is minimized for a given sensitivity to simultaneously satisfy adversary's conflicting goals for the problem of adversarial classification under Gaussian DP-mechanism. 
Also note that the additional factor $2\pi \mathrm{e}$ appears in our bounds as opposed to the original rate-distortion function and the corresponding lower bound on squared-error distortion due to the choice of the query function that aggregates the entire dataset and returns an output of size 1. 
\end{remark}
\subsection{A Statistical Threshold to Avoid Detection- First-Order Statistics \label{sec:statistic}}

Next, we present a statistical trade-off between the security of the Gaussian mechanism and the adversary's advantage.
\begin{theorem} \label{theorem:k_Gauss}
The adversary avoids being correctly detected by the defender with the largest possible power of the test $\bar{\beta}=1-\beta$ and the best critical region of size $\alpha$ for positive bias, if the following inequality holds
\begin{equation} \label{eq:theo11}
\Delta \mu \leq \left(Q^{-1}(\alpha) -Q^{-1}(\bar{\beta})\right) \sigma_z
\end{equation} where $Q(.)$ denotes the Gaussian Q-function defined as $\Pr[S>s]$ and for $\sigma_z= \frac{\sqrt{2} \cdot s \cdot .5 \cdot \log (1.25/\delta)}{\epsilon}$. By analogy, for negative bias, we have
\begin{equation}\label{eq:theo12}
\Delta \mu \geq \left(Q^{-1}(\bar{\alpha}) -Q^{-1}(\beta)\right) \sigma_Z
\end{equation} where $\bar{\alpha}=1-\alpha$.
\end{theorem}
\begin{proof}
Likelihood ratio function $\Lambda$ to choose between $Y-\sum_i^n X_i$ and $Y-\sum_i^n X_i - X_a$ results in
 $z> \tilde{k}$ where $\tilde{k}= \frac{\sigma_z^2 \log k}{\Delta \mu}+\frac{\mu_1+\mu_0}{2}$ by setting $p_0$ and $p_1$ as Gaussian distributions with respective location parameters $\mu_0$ and $\mu_1$ and the mutual scale parameter $\sigma_z$. Probability of rejecting $H_0$ in case of an attack is derived using this condition as
\begin{equation} \label{eq:alpha}
\alpha=
\begin{cases}
Q\left(\frac{\sigma_z \log k}{\Delta \mu}+\frac{\Delta \mu}{2 \sigma_z}\right)\;\Delta \mu >0,\\
1- Q \left(\frac{\sigma_z \log k}{\Delta \mu}+\Delta \mu/(2 \sigma_z)\right)
\end{cases}
\end{equation}where $Q(.)$ denotes the Gaussian Q-function defined as $\Pr[T>t]$ for standard Gaussian random variables. The threshold of the critical region $k$ for $\Delta \mu>0$ is obtained as a function of the probability of false-alarm as
$k={\exp\left\{\frac{\Delta \mu}{\sigma_z} \left(Q^{-1}(\alpha)-\Delta \mu/2\sigma_z\right)\right\}}$. The second threshold for negative bias can be obtained similarly. The defender fails to detect the attack if $Y< k+q(\mathbf{X})$, where $q(.)$ is the noiseless query output. By analogy, for $\Delta \mu<0$, the attack is not detected if the DP output exceeds $\bar{k}+q(\mathbf{X})$ where $\bar{k}=\exp\left\{\frac{\Delta \mu}{\sigma_z} \left(Q^{-1}(\bar{\alpha})-\frac{\Delta \mu}{2\sigma_z}\right)\right\}$.
The power of the test for both cases is obtained as follows
\begin{equation}\label{eq:beta}
\bar{\beta}= 
\begin{cases}
Q\left(Q^{-1}(\alpha)-\Delta \mu/\sigma_z\right),\; \mathrm{for}\;\Delta \mu >0,\\
1- Q\left(Q^{-1}(\bar{\alpha})-\Delta \mu/\sigma_z\right)\; \mathrm{for}\;\Delta \mu <0.
\end{cases}
\end{equation} Rewriting (\ref{eq:alpha}) and (\ref{eq:beta}) results in (\ref{eq:theo11}) and (\ref{eq:theo12}). \end{proof} 
Numerical evaluation results of Theorem \ref{theorem:k_Gauss} are presented in Section \ref{sec:numeric}.
\section{Chernoff DP \label{sec:chernoff}}

In the classical approach, the best error exponent in hypothesis testing for choosing between two probability distributions is the Kullback-Leibler divergence between these two distributions due to Stein's lemma \cite{r14}. In the Bayesian setting, however, assigning prior probabilities to each of the hypotheses in a binary hypothesis testing problem minimizes the best error exponent when the weighted sum probability of error, i.e. $\pi= a \alpha + b \beta$ for $b=1-a$ and $a\in (0,1)$ which corresponds to the \textit{Chernoff information/divergence}.
The Chernoff information between two probability distributions $f_0$ and $f_1$ with prior probabilities $a$ and $b$ is defined as
\begin{equation} \label{def_chernoff}
C_a(f_0||f_1)=\log \int_x f_0(x)^a f_1^{b}(x) dx
\end{equation} The Renyi divergence denoted $D_{a}(f_0||f_1)$ between two Gaussian distributions with parameters $\mathcal{N}(\mu_0, \sigma_0^2)$ and $\mathcal{N}(\mu_1, \sigma_1^2)$ is given in \cite{GAL13} by 
\begin{align}
D_{a}(f_0||f_1)&= \ln \frac{\sigma_1}{\sigma_0}+\frac{1}{2(a-1)}\ln \left(\frac{\sigma_1^2}{(\sigma^2)^*_a}\right) \\
&+\frac{1}{2}\frac{a(\mu_0-\mu_1)^2}{(\sigma^2)^*_a} 
\end{align} where $(\sigma^2)^*_a= a \sigma^2_1 +b \sigma^2_0$. Using the following relation between Chernoff information and Renyi divergence $D_{a}(f_0||f_1)=\frac{1}{1-a} C_a(f_0||f_1)$, we obtain the Gaussian univariate Chernoff information 
with different standard deviations $\sigma_i$ for $i=0,1$ as follows.
\begin{equation}
C(f_0||f_1)= b \ln \frac{\sigma_1}{\sigma_0}+\frac{1}{2}\ln \frac{\sigma^2_1}{a \sigma^2_1 +b \sigma^2_0}+\frac{a \cdot b}{2}\frac{(\mu_0-\mu_1)^2}{a \sigma^2_1 +b \sigma^2_0}. \notag
\end{equation} On the other hand, KL divergence between two Gaussian distributions denoted $D_{KL}(f_0||f_1)$ is derived as $\log\left(\frac{\sigma_1}{\sigma_0}\right)+\frac{1}{2}\frac{\sigma_0^2}{\sigma_1^2}+\frac{(\mu_1-\mu_0)^2}{2\sigma_1^2}-\frac{1}{2}$. 

The next definition provides an adaptation of Chernoff information to quantify DP guarantee as a stronger alternative to KL-DP of Definition \ref{def:KL-DP} and $(\epsilon, \delta)-$DP for Gaussian mechanisms. We apply this to our problem setting for adversarial classification under Gaussian mechanisms, where the query output before and after the attack are $\sum_i^n X_i$ and $\sum_i^n X_i+X_a$, respectively. The corresponding distributions are considered as the DP noise with and without the induced value of $X_a$ by the attacker as in our original hypothesis testing problem in (\ref{eq:ht}) in Section \ref{subsec:model_first}.
\begin{definition}[Chernoff DP]\label{def:chernoff_dp}
For a randomized mechanism $P_{Y|X}$ guarantees $\epsilon-$ Chernoff-DP, if the following inequality holds for all its neighboring datasets $x$ and $\tilde{x}$
\begin{equation} \label{eq:C_dp}
C_a(P_{Y|X=x}||P_{Y|X=\tilde{x}}) \leq \exp(\epsilon)
\end{equation} where $C_a(.||.)$ is defined by (\ref{def_chernoff}).
\end{definition} 
\cite[Theorem 1]{CY16} proves that KL-DP defined in Definition \ref{def:KL-DP} is a stronger privacy metric than $(\epsilon, \delta)-$DP that is achieved by Gaussian mechanism. Accordingly, the following chain of inequalities are proven to hold for various definitions of DP
\begin{equation} 
\epsilon-\mathrm{DP} \overset{a}{\succeq} \mathrm{KL-DP} \overset{b}{\succeq} \mathrm{MI-DP}\overset{c}{\succeq} \delta-\mathrm{DP} \overset{d}{=} (\epsilon, \delta)-\mathrm{DP} \notag
\end{equation} where MI-DP refers to the mutual information DP defined by $\underset{i, P_{X^n}}{\sup} I(X_i;Y|X^{-i}) \leq \epsilon \;\mathrm{nats}$ for a dataset $X^n=\{X1, \cdots, X_n\}$ with the corresponding output $Y$ according to the randomized mechanism represented by $P_{Y|X^n}$ where $X^{-i}$ denotes the  dataset entries excluding $X_i$. $\delta-$DP represents the case when $\epsilon=0$ in $(\epsilon,\delta)-$ DP.

Chernoff information based definition of DP is a \textit{\textbf{stronger privacy metric}} than KL-DP, and thus $(\epsilon, \delta)-$DP for the Gaussian mechanism due to prior probabilities. Such a comparison is presented numerically in Figure \ref{fig:comp_KL_Chernoff}. Numerical evaluation also supports the same conclusion. 
For the special case of equal standard deviation of both distributions, Chernoff information $C(f_0||f_1)$ is exactly $a \cdot b \cdot D_{KL}(f_0||f_1)$.
\section{Numerical Evaluations \label{sec:numeric}}

\begin{figure}[h!]
 \centering
 \includegraphics[width=1\linewidth]{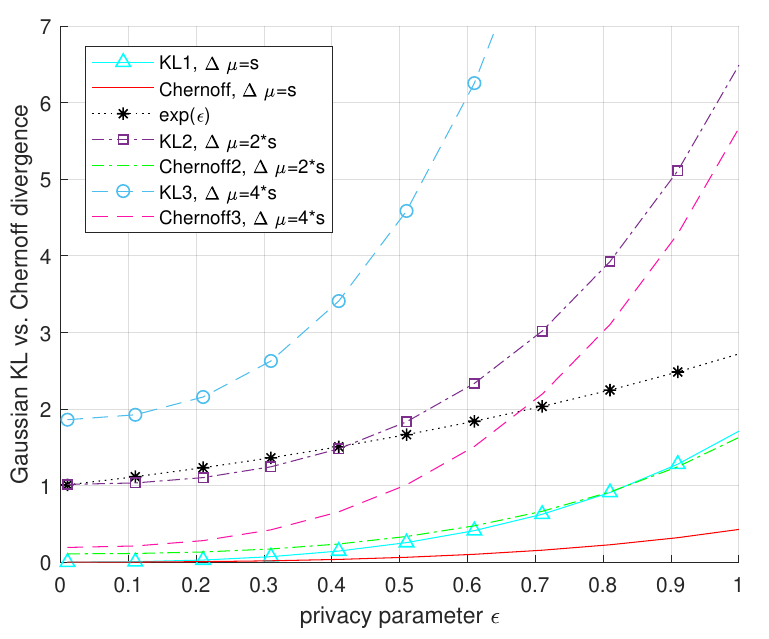}
\caption{KL-DP vs. Chernoff DP for various levels of privacy budget with global sensitivity $s=4$}
\label{fig:comp_KL_Chernoff}
 \end{figure}

Figure \ref{fig:comp_KL_Chernoff} depicts Chernoff DP and KL-DP for various levels of privacy and the impact of the attack which were set as a function of the global sensitivity. Accordingly, the attack is compared to the privacy constrained of Definition \ref{def:chernoff_dp} that is referred as the upper bound in the legend. Due to prior probabilities, Chernoff information is tighter than KL divergence consequently, it provides a more strict privacy constraint. Figure \ref{fig:comp_KL_Chernoff} confirms that increasing the impact of the attack as a function of the sensitivity closes the gap with the upper bound for Chernoff-DP. Additionally, the KL-DP does not violate the upper bound of the privacy budget only in the high privacy regime (when $\epsilon$ is small) for the cases of $\Delta \mu=2\cdot s $ and $\Delta \mu=4 \cdot s$.  

\begin{figure}[!h]
 \centering
 \includegraphics[width=1\linewidth]{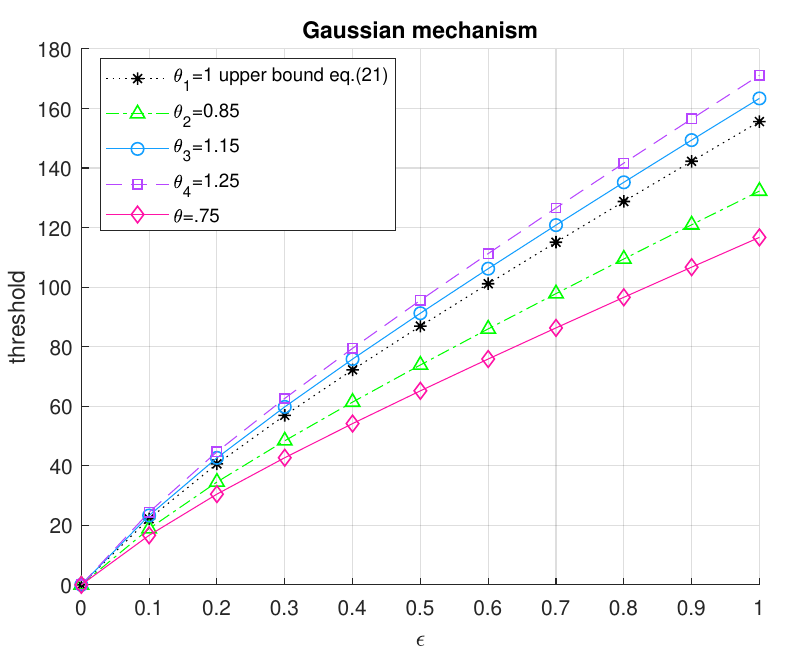}
\caption{Numerical comparison of the upper bound (\ref{upper_bound_attack}) with the thresholds of the critical region}
\label{fig:advclass_inf}
 \end{figure}

The upper bound (\ref{upper_bound_attack}) on the additional data's variance presented in Corollary \ref{cor:upper_bound}, is tested for two opposing hypothesis in (\ref{eq:ht}) and the corresponding thresholds of the critical region (to be compared to the chi-square table values) are depicted in Figure \ref{fig:advclass_inf}. Here the null hypothesis that states that the defender fails to detect the attack corresponds to the case where $\sigma^2_{X_a}$ respects the upper bound (\ref{upper_bound_attack}) whereas the alternative hypothesis claims the variance of $X_a$ exceeds the proposed bound by factors stated in the legend of the figure. Increasing the privacy budget also increases the threshold and $\theta \sigma^2_{X_a}$ violates the upper bound for $\theta>1$. This is consistent with Figure \ref{fig:advclass_inf}.

\begin{figure}[ht!]
 \centering
 \includegraphics[width=1.05\linewidth]{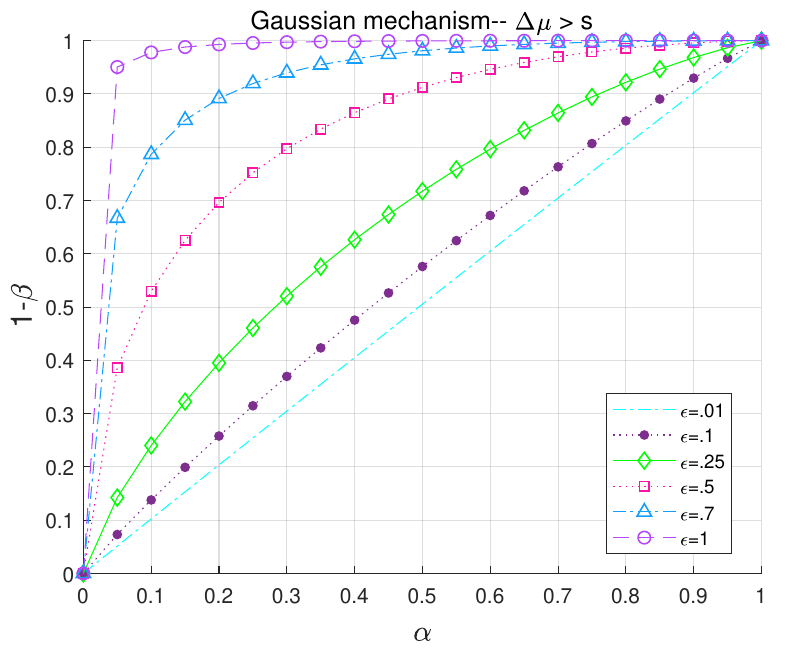}
\caption{Eqs. (\ref{eq:alpha}) and (\ref{eq:beta}) for various values of $\epsilon$, $\Delta \mu>0, \; \delta=\epsilon/20$ and $\Delta \mu >s$.} 
\label{fig:means_gauss_onesided1}
 \end{figure}

\begin{figure}[ht!]
 \centering
 \includegraphics[width=1.05\linewidth]{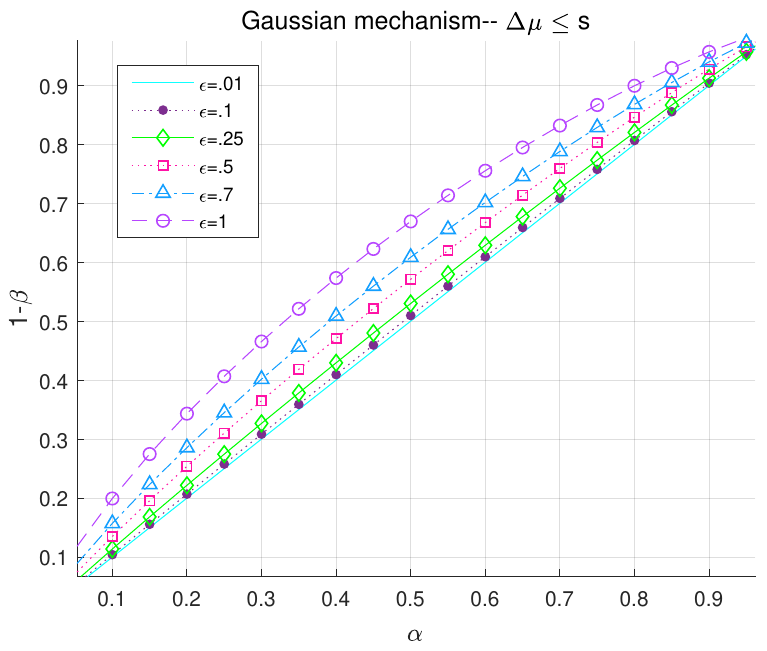}
 \caption{Eqs. (\ref{eq:alpha}) and (\ref{eq:beta}) for various values of $\epsilon$, $\Delta \mu>0, \;\delta=\epsilon/20$ and $\Delta \mu \leq s$.} 
\label{fig:means_gauss_onesided2}
 \end{figure}

Figures \ref{fig:means_gauss_onesided1} and \ref{fig:means_gauss_onesided2} presents ROC curves computed using the threshold of (\ref{eq:theo11}) for adversarial classification under Gaussian DP for three different scenarios where the impact of the attack is greater than, equal to and less than the $L_2$ norm global sensitivity (in this order) for various levels of privacy budget. We observe that in the low privacy regime (i.e. when $\epsilon$ is large) the accuracy of the test is high which comes at the expense of the privacy guarantee since as the privacy budget is decreased (higher privacy) the test is no longer accurate and the adversary cannot be correctly detected with high probability. Another observation can be made based on the effect of the relationship between the attack and sensitivity. Unsurprisingly, increasing the bias $\Delta \mu$ as opposed to $s$ also increases the probability of correctly detecting the attacker.  

\section{Conclusion \label{sec:conc}}
We established statistical and information-theoretic trade-offs between the security of the Gaussian DP-mechanism and the adversary's advantage who aims to trick the classifier that detects anomalies. Accordingly, we determined a statistical threshold that offsets the DP-mechanism's privacy budget against the impact of the adversary's attack to remain undetected and introduced the privacy-distortion function which we used for bounding the impact of the adversary's modification on the original data.
We introduced Chernoff DP and its application to adversarial classification which turned out to be a stronger privacy metric than KL-DP and $(\epsilon, \delta)-$DP for the Gaussian mechanism.

                                      
\bibliography{paper_adv_dp}

\begin{thebibliography}{10}
\providecommand{\url}[1]{#1}
\csname url@samestyle\endcsname
\providecommand{\newblock}{\relax}
\providecommand{\bibinfo}[2]{#2}
\providecommand{\BIBentrySTDinterwordspacing}{\spaceskip=0pt\relax}
\providecommand{\BIBentryALTinterwordstretchfactor}{4}
\providecommand{\BIBentryALTinterwordspacing}{\spaceskip=\fontdimen2\font plus
\BIBentryALTinterwordstretchfactor\fontdimen3\font minus
  \fontdimen4\font\relax}
\providecommand{\BIBforeignlanguage}[2]{{%
\expandafter\ifx\csname l@#1\endcsname\relax
\typeout{** WARNING: IEEEtran.bst: No hyphenation pattern has been}%
\typeout{** loaded for the language `#1'. Using the pattern for}%
\typeout{** the default language instead.}%
\else
\language=\csname l@#1\endcsname
\fi
#2}}
\providecommand{\BIBdecl}{\relax}
\BIBdecl

\bibitem{D06}
C.~{D}work, ``Differential privacy,'' in \emph{Automata, Languages and
  Programming}, 2006, pp. 1--12.

\bibitem{r1}
J.~Giraldo, A.~A. Cardenas, M.~Kantarcioglu, and J.~Katz, ``{A}dversarial
  {C}lassification {U}nder {D}ifferential {P}rivacy,'' in \emph{{NDSS} 2020,
  {N}etwork and {D}istributed {S}ystems {S}ecurity {S}ymposium, {S}an {D}iego,
  {CA}, {USA}}, Feb. 2020.

\bibitem{r12}
M.~{L}ecuyer, V.~{A}tlidakis, R.~{G}eambasu, D.~{H}su, and S.~{J}ana,
  ``Certified robustness to adversarial examples with differential privacy,''
  in \emph{IEEE Symposium on Security and Privacy, San Francisco CA, USA}, May
  2019, pp. 1054--1067.

\bibitem{UO21}
A.~\"{U}nsal and M.~\"{O}nen, ``A {S}tatistical {T}hreshold for {A}dversarial
  {C}lassification in {L}aplace {M}echanisms,'' in \emph{IEEE Information
  Theory Workshop 2021}, Oct. 2021.

\bibitem{WYZ16}
W.~{W}ang, L.~{Y}ing, and J.~{Z}hang, ``On the relation between
  identifiability, differential privacy and mutual information privacy,''
  \emph{IEEE Transactions on Information Theory}, vol.~62, pp. 5018--5029, Sep.
  2016.

\bibitem{SS14}
A.~{S}arwate and L.~{S}ankar, ``A rate-distortion perspective on local
  differential privacy,'' in \emph{Fiftieth Annual Allerton Conference}, Oct.
  2014, pp. 903--908.

\bibitem{CF12}
F.~du~{P}in Calmon and N.~{F}awaz, ``Privacy against statistical inference,''
  in \emph{Fiftieth Annual Allerton Conference}, Oct. 2012, pp. 1401--1408.

\bibitem{PG21}
A.~{P}astore and M.~{G}astpar, ``Locally differentially private randomized
  response for discrete distribution learning,'' \emph{Journal on Machine
  Learning Research}, vol.~22, pp. 1--56, Jul. 2021.

\bibitem{CY16}
P.~{C}uff and L.~{Y}u, ``{D}ifferential {P}rivacy as a {M}utual {I}nformation
  {C}onstraint,'' in \emph{{CCS} 2016, {V}ienna, {A}ustria}, Oct. 2016.

\bibitem{DR05}
C.~{D}work and A.~{R}oth, ``The {A}lgorithmic {F}oundations of {D}ifferential
  {P}rivacy,'' \emph{Foundations and Trends in Theoretical Computer Science
  2014}, vol.~9, pp. 211--407, 2014.

\bibitem{DMNA06}
C.~{D}work, F.~{M}cSherry, K.~{N}issim, and A.~{S}mith, ``Calibrating {N}oise
  to {S}ensitivity in {P}rivate {D}ata {A}nalysis,'' in \emph{Theory of
  Cryptography Conference}, 2006, pp. 265--284.

\bibitem{r3}
C.~{L}iu, X.~{H}e, T.~{C}hanyaswad, S.~{W}ang, and P.~{M}ittal, ``Investigating
  {S}tatistical {P}rivacy {F}rameworks from the {P}erspective of {H}ypothesis
  {T}esting,'' in \emph{{PETS} 2019 Proceedings on Privacy Enhancing
  Technologies}, 2019, pp. 233--254.

\bibitem{r14}
T.~Cover and J.~A. Thomas, \emph{Elements of Information Theory}.\hskip 1em
  plus 0.5em minus 0.4em\relax Wiley Series in Telecommunications, 1991.

\bibitem{GAL13}
M.~{G}il, F.~{A}lajaji, and T.~{L}inder, ``Renyi divergence measures for
  commonly used univariate continuous distributions,'' \emph{Information
  Sciences}, vol. 249, Nov. 2013.

\end{thebibliography}
\bibliographystyle{IEEEtran}


\end{document}